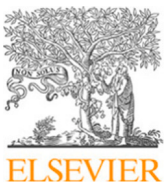
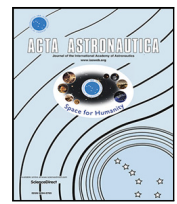

Research paper

# Projections of Earth's technosphere: Luminosity and mass as limits to growth

Jacob Haqq-Misra [a],*, Clément Vidal [b], George Profitiliotis [a]

[a] *Blue Marble Space Institute of Science, Seattle, WA, USA*
[b] *Center Leo Apostel, Vrije Universiteit Brussel, Brussels, Belgium*

ABSTRACT

Earth remains the only known example of a planet with technology, and future projections of Earth's trajectory provide a basis and motivation for approaching the search for extraterrestrial technospheres. Conventional approaches toward projecting Earth's technosphere include applications of the Kardashev scale, which suggest the possibility that energy-intensive civilizations may expand to harness the entire energy output available to their planet, host star, or even the entire galaxy. In this study, we argue that the Kardashev scale is better understood as a "luminosity limit" that describes the maximum capacity for a civilization to harvest luminous stellar energy across a given spatial domain, and we note that thermodynamic efficiency will always keep a luminosity-limited technosphere from actually reaching this theoretical limit. We suggest the possibility that an advanced technosphere might evolve beyond this luminosity limit to draw its energy directly from harvesting stellar mass, and we also discuss possible trajectories that could exist between Earth today and such hypothetical "stellivores." We develop a framework to describe trajectories for long-lived technospheres that optimize their growth strategies between exploration and exploitation, unlike Earth today. We note that analyses of compact accreting stars could provide ways to test the stellivore hypothesis, and we more broadly suggest an expansion of technosignature search strategies beyond those that reside exactly at the luminosity limit.

## 1. Introduction

The search for technosignatures depends on making meaningful projections of Earth's future in order to inform our thinking and planning. Projections of the future in the context of the search for technosignatures often assume continuous exponential growth and expansion across the galaxy, which is based on a model that was developed by Kardashev [1]. The original Kardashev scale included a Type I civilization that has harnessed all the power available to its planet, a Type II civilization that has harnessed all available stellar power, and a Type III civilization that has harnessed the power of all stars in a galaxy. The Kardashev scale remains an important reference point for discussing possible observational characteristics and motivating actual searches for such energy-intensive civilizations.

The idea of continuous growth that underlies the Kardashev scale contains two embedded assumptions: (1) growth in energy use and (2) growth in spatial scale. Kardashev assumed that human civilization would continue increasing its energy consumption at a 1% annual growth rate, which would eventually reach a point at which all sunlight falling on the planet (Type I) or all sunlight emitted from the star (Type II) would be required. The original intent of Kardashev [1] was to place constraints on the transmitting power for interstellar communication, which was focused on growth in energy use that corresponded to the different spatial scales. However, many subsequent analyses have assumed that this projected rate in annual energy use is correlated with an increase in the spatial domain available to the civilization; or in other words, a civilization with exponentially increasing energy needs is assumed to also proportionally expand across the galaxy. These two assumptions are taken together in many applications of the Kardashev scale [see, e.g., 2, for an overview], and the analysis by Baghram [3] even explicitly defines a "space exploration distance" that is tightly coupled to energy consumption according to the Kardashev scale. However, it is not necessarily evident that future changes in the energy use or spatial domain of Earth (or an extraterrestrial technosphere) must follow this trajectory.

The idea of continuous growth underlying the Kardashev scale in some ways reflects the great acceleration [4] that was observed from the 1950s onward, so the idea that growth would necessarily continue unbounded and at an exponential pace may reflect an anthropocentric bias of these historical conditions. For example, the projections by Von Hoerner [5] assumed a 7% growth rate that was indicated by the data at the time, although more recent analyses of energy use trends show this has slowed to around 2% [e.g., 6]. Such growth rates in population and energy consumption may continue to reduce, even reaching a point of zero growth. Likewise, technological developments (such as generalized artificial intelligence) could conceivably show such rapid increases in efficiency that enable the ability to evolve without depending on an exponentially increasing supply of energy. But the energetic interpretation of the Kardashev scale may also be more universally applicable: all living things and all complex systems need free energy to function, whether they are cells, plants, animals, technological artifacts, cities, or spacefaring civilizations. The availability of free energy also can be utilized to enable a vast range of behaviors or motivations that need not be anthropocentric, so in this sense the Kardashev scale is a generalized approach toward thinking about growth and expansion. This universality of energy is arguably one reason for the continued relevance of the Kardashev scale 60 years after having been introduced [see., e.g.,2].

**Table 1**
Energy use and spatial domain for present-day Earth, the extended Kardashev scale [7], and projections of Earth's 1000-year future [8] sorted by ascending annual energy use.

| Scenario | Annual Energy Use (J) | Domain radius (m) | Domain description |
|---|---|---|---|
| *Present-day* | | | |
| Earth | $6 \times 10^{20}$ | $6.4 \times 10^{6}$ | Earth |
| *Extended Kardashev Scale* | | | |
| Type I | $3 \times 10^{23}$ | $6.4 \times 10^{6}$ | Planet |
| Type II | $3 \times 10^{33}$ | $3.5 \times 10^{13}$ | Star |
| Type III | $3 \times 10^{43}$ | $2 \times 10^{20}$ | Galaxy |
| Type IV | $3 \times 10^{53}$ | $10^{27}$ | Universe |
| *1000-year Projections* | | | |
| Living with the Land (S4) | $3 \times 10^{19}$ | $6.4 \times 10^{6}$ | Earth |
| Restoration (S7) | $3 \times 10^{20}$ | $6.4 \times 10^{6}$ | Earth |
| Golden Age (S3) | $8 \times 10^{20}$ | $4.5 \times 10^{12}$ | Venus to Neptune |
| Ouroboros (S8) | $2 \times 10^{21}$ | $3.8 \times 10^{8}$ | Earth to Moon |
| Big Brother is Watching (S1) | $2 \times 10^{21}$ | $4.5 \times 10^{12}$ | Venus to Neptune |
| Out of Eden (S10) | $2 \times 10^{21}$ | $2 \times 10^{13}$ | Mercury to Heliopause |
| Transhumanism (S5) | $3 \times 10^{21}$ | $7.4 \times 10^{12}$ | Venus to Kuiper Belt |
| Wild West (S2) | $8 \times 10^{21}$ | $3.4 \times 10^{11}$ | Earth to Asteroid Belt |
| Sword of Damocles (S6) | $2 \times 10^{22}$ | $7.4 \times 10^{12}$ | Venus to Kuiper Belt |
| Deus Ex Machina (S9) | $10^{25}$ | $2 \times 10^{13}$ | Mercury to Heliopause |

This paper suggests that the Kardashev scale is best interpreted as a "luminosity limit", which describes the maximum capacity for harnessing luminous stellar energy across a spatial domain. We retain the assumption that stars are the primary source of energy available to long-lived technospheres. We suggest that many plausible technosphere scenarios could exist below the luminosity limit. We also discuss potential cases in which other methods of harvesting stellar energy could exceed the luminosity limit and approach a limit based on the availability of mass in a system.

## 2. The luminosity limit

The basic idea underlying our approach is that a technosphere, as well as a biosphere, needs both energy and resources. The Kardashev scale links both of these requirements, but we present a framework to more explicitly show the relationship between these variables. We consider several technosphere scenarios in terms of the annual energy use (in Joules) and the spatial domain radius (in meters). The spatial domain describes the largest physical volume in which a technosphere can obtain resources; it is not necessarily required that this entire volume is inhabited or fully utilized, but only that it is utilizable. We summarize all the scenarios considered in Table 1 and Fig. 1.

The values for energy use for the Kardashev scale follow the extended classification scheme developed by Gray [7], which includes a Type IV civilization that has harnessed the power of the entire universe. The domain radius for the Type I, II, III, and IV scenarios is the maximum spatial extent required to achieve the corresponding annual energy use from harnessing stellar luminosity. The Type I spatial domain is the radius of Earth. The Type II spatial domain is an upper limit for the termination shock where the solar wind meets the interstellar medium. The Type III spatial domain is the radius of the Milky Way galaxy. The Type IV spatial domain is the approximate size of the observable universe. This enables the Kardashev scale to be expressed as a function of domain radius and annual energy use, which is shown in Fig. 1 as the "luminosity limit". (This can be compared with the coupling between the Kardashev energy consumption levels and the "space exploration distance" defined by Baghram [3].) We can fit a line to the luminosity limit to find an approximate functional relationship between the maximum annual energy use available from stellar luminosity, $E_{max}^{lum}$, and the spatial domain radius, $R$:

$$E_{max}^{lum} \approx R^{1.5} \cdot 10^{13.4} \text{ J m}^{-1}. \tag{1}$$

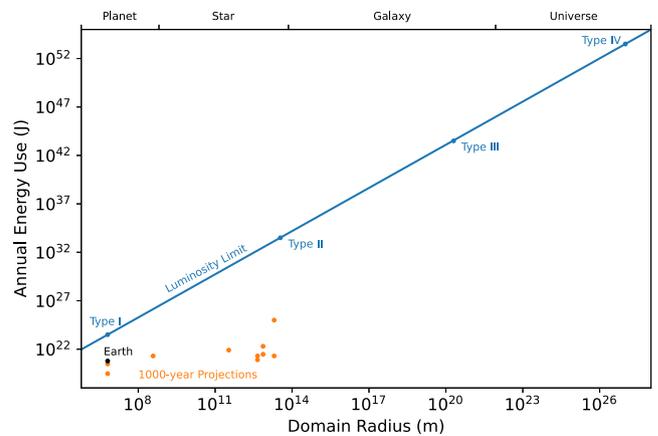

**Fig. 1.** Log–log plot of annual energy use versus spatial domain radius for a technosphere. The luminosity limit is defined by the energy use and spatial domain associated with the extended Kardashev scale [7]. Earth today and projections of Earth's 1000-year future [8] all fall below the luminosity limit. Values are provided in Table 1.

This expression is calculated by performing a linear fit between the logarithmic values of the domain radius and annual energy use for Type I and Type IV civilizations, as listed on Table 1. Eq. (1) is not a prescription for how a technosphere must grow but instead is an upper limit to the energy consumption available within a spatial domain. Rather than a prediction of future growth, the luminosity limit can be an instructive tool for thinking about possible future trajectories.

Earth today falls below the luminosity limit because the total energy use of human civilization is less than the total energy available from sunlight on Earth's surface. The spatial domain available to human civilization is the entire planet, which reflects the ability of human civilization to travel to and utilize almost any resource on Earth, if desired. This resource domain could extend in the future, for example if asteroid mining or Mars settlement become viable. Such possibilities can inform the development of scenarios in which the technosphere expands beyond Earth.

We consider the set of ten self-consistent projections of Earth's technosphere 1000 years into the future developed by Haqq-Misra et al. [8], which are all within the lower left quadrant of Fig. 1. These ten scenarios were developed by conducting a morphological analysis of the political, economic, social, and technological scenario space for Earth's 1000-year future, performing a cross-consistency analysis to eliminate inconsistent scenarios, and then applying a novel worldbuilding pipeline to identify the observable properties of the technosphere. The spatial domain of the scenarios ranges from Earth alone to the entire solar system, and the range of energy use includes scenarios both less than and greater than Earth's energy use today. All ten scenarios fall below the luminosity limit.

This set of scenarios includes only one that has exceeded the Type I threshold for annual energy use; however, this case still falls below the luminosity limit because the domain radius is much larger than the planet. The domain radius for this and several other cases in the ten scenarios is much closer to the Type II domain radius, even though energy use is far below the Type II threshold for annual energy use. In this set of scenarios, annual energy use increases at a much slower rate with domain radius when compared to the luminosity limit. This suggests that energy use and expansion may not be as tightly coupled as assumed in the Kardashev [1] scale, and plausible projections of Earth's technosphere might be expected to fall below the luminosity limit rather than to follow it.

### 3. Stellivores: Beyond the luminosity limit

The luminosity limit is a hard limit for any technosphere that is based on harnessing stellar energy within a spatial domain, but such a limit would not necessarily apply to a technosphere that can extract even larger quantities of energy from a star. No such cases emerged from the scenario modeling by Haqq-Misra et al. [8], but the parameter space shown in Fig. 1 raises the possibility of technospheres that might exist above the luminosity limit.

A hypothetical idea by Vidal [9] is that some accreting binary stars might actually be living organisms or technological systems, with one compact object "feeding" on its companion star to sustain metabolism. Some "stellivores" may also have higher velocities with trajectories that are directed toward nearby stars for continued feeding [10]. In such a case a stellivore would be observed as a binary system not accreting, but ejecting material out of its gravitational well, in order to generate thrust to travel toward another star [11]. Observationally testing the stellivore hypothesis is challenging, but possible, and Vidal [12] suggested an approach based on characterizing the energy rate density of such systems. In principle, the idea that some accreting binary systems could be stellivores is a possibility worth considering as a potential living organism or technological system that could exceed the luminosity limit. Examples of possible stellivore candidates are listed in Table 2 and shown on Fig. 2. These values for accreting white dwarfs, neutron stars, and black holes are average quantities that were tabulated by Vidal [12].

The spatial domain for stellivores takes two separate cases. First is the case of "stationary stellivores" that do not necessarily move toward new feeding targets, in which case the spatial domain is the orbital separation between the two stars. The stellivore hypothesis proposes three levels of density: white-dwarf, neutron-star and black hole density. The spatial domain for stationary stellivores is chosen as the typical orbital separation for cataclysmic variables (white dwarfs accreting) and low-mass X-ray binaries (neutron stars or black holes accreting), which is about 1–2 solar radii or about $10^9$ meters [13]. Second is the case of "traveling stellivores" that move over a larger region of space to feed on new stars as needed. In this case, the spatial domain is a small globular cluster with radius of 10 light years ($\sim 10^{17}$ meters), which would provide numerous opportunities for close stellar encounters.

All stationary stellivores in Fig. 2 exceed the luminosity limit. For traveling stellivores in a globular cluster, the accreting neutron star and black hole cases both exceed the luminosity limit, while the accreting white dwarf case is very close to the luminosity limit. The idea that a living or technological system could extract energy from a star through non-luminous processes, such as accretion, suggests possibilities for exceeding the luminosity limit.

Other possibilities can be imagined by thinking about the evolutionary trajectory of stellivores or other technological processes that could operate above the luminosity limit. A pre-stellivore phase could involve planet-like accretion that occurs very close to the host star and at a lower energy consumption rate than stationary stellivores, which is above the luminosity limit with the spatial domain approximately equal to the radius of the Sun. Another possibility is the technology of "star-lifting", which would involve removing sufficient mass to prevent the host star from evolving into a red giant [14]. Star-lifting would require energy expenditures above the luminosity limit, if the spatial domain is taken as the Earth–Sun separation. These hypotheses are worth exploring further in order to make predictions about civilizations interacting with their star, which might be observable with present or near-future astronomical facilities.

Even more exotic sources could provide vast and longer-term energy reservoirs than the consumption of stellar matter. For example, some speculative research has suggested the potential feasibility of harvesting zero-point energy (i.e, quantum background energy) for enabling high-velocity spaceflight [15]. Some of the initial technical challenges associated with this technology may be tractable [16], although the actual construction of any apparatus for realistically harnessing zero-point energy remains a future challenge for human civilization. If any extraterrestrial civilization were able to harvest zero-point energy, then it could provide a near limitless source of energy orders of magnitude beyond stellar luminosity or mass limits. A study by Wright et al. [17] suggested that a galaxy-spanning civilization that harnesses zero-point (or other non-stellar) sources of energy would necessarily expel large quantities of thermal waste energy, so searching for the presence of such mid-infrared excesses from galactic sources is one way to constrain the presence of such Type III civilizations. The data analysis by Wright et al. [18] found no such sources in data from the Wide-field Infrared Survey Explorer (WISE) missions, which implies that no Type III civilization is harnessing such exotic physics at high effective temperatures. Either such physics is impossible, or else any Type III civilization that does exist has found a way to expel thermal waste energy at much lower temperatures (i.e., at longer wavelengths than the WISE survey can detect) or through other means (i.e., expelling waste energy as neutrinos). Such exotic possibilities are worth considering but difficult to quantify, so they will be only mentioned here as a theoretical possibility for exceeding the luminosity limit.

### 4. Thermodynamic efficiency

A major transformation would be required for an Earth-like technosphere to exceed the luminosity limit, such as in the example of stellivores. We cannot easily predict the trajectory of such a technosphere, but we can gain some insight into the necessary transformation by applying a systems thermodynamic framework to this problem.

One way of framing the second law of thermodynamics is using the concept of *exergy*, where exergy refers to the work obtainable within a system. Using this concept, the second law of thermodynamics states that some "exergy loss" (i.e., entropy production) will always occur in a system, so that there can never be perfect efficiency between exergy input to the system (i.e., the available useful energy) and exergy output from the system (i.e., products and emissions). The ratio between the exergy outflow and exergy inflow is the "exergetic efficiency" for a system [see, e.g., 19].

**Table 2**
Energy use and spatial domain for several stellivore scenarios. Note that Vidal [12] lists values as energy rate density for an associated stellar mass, which have been converted here into units of energy per year.

| Scenario | Annual energy use (J) | Domain radius (m) | Domain description |
| --- | --- | --- | --- |
| *Stationary Stellivores* | | | |
| Accreting White Dwarfs | $2.4 \times 10^{38}$ | $10^{9}$ | Solar diameter |
| Accreting Neutron Stars | $2.8 \times 10^{40}$ | $10^{9}$ | Solar diameter |
| Accreting Black Holes | $2.2 \times 10^{41}$ | $10^{9}$ | Solar diameter |
| *Traveling Stellivores* | | | |
| Accreting White Dwarfs | $2.4 \times 10^{38}$ | $10^{17}$ | Globular Cluster |
| Accreting Neutron Stars | $2.8 \times 10^{40}$ | $10^{17}$ | Globular Cluster |
| Accreting Black Holes | $2.2 \times 10^{41}$ | $10^{17}$ | Globular Cluster |
| *Star-lifting* | | | |
| Lower Limit for Mass Removal | $2.4 \times 10^{31}$ | $10^{11}$ | Earth to Sun |
| Upper Limit for Mass Removal | $2.4 \times 10^{33}$ | $10^{11}$ | Earth to Sun |
| *Planet-like Accretion* | | | |
| Close-orbiting pre-stellivore | $\sim 10^{32}$ | $\sim 10^{9}$ | Radius of Sun |

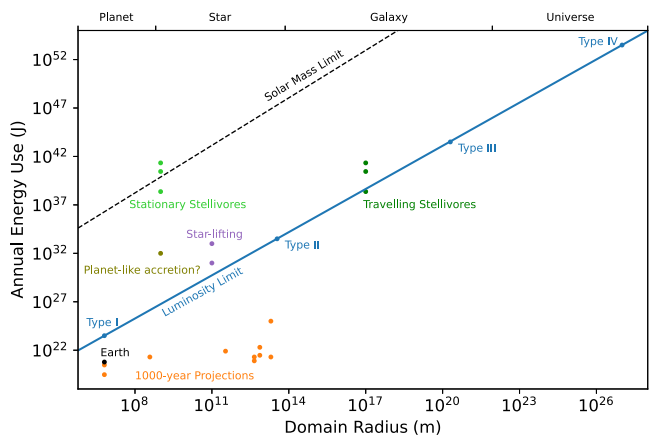

**Fig. 2.** Log–log plot of annual energy use versus spatial domain radius for hypothetical star-eating "stellivores" [9] that extract energy from stars through accretion and could potentially exceed the luminosity limit. The "star-lifting" cases describe a technosphere that is prolonging the lifetime of its host star by removing stellar mass [14]. Values are provided in Table 2. The solar mass limit is defined by the maximum annual energy use that could be sustained from harvesting mass from a sun-like star.

The luminosity limit describes the maximum annual energy use available from stellar radiation, $S$, which corresponds to the theoretical maximum exergy input to a technosphere based on stellar radiation alone. Because of the second law of thermodynamics, a luminosity-limited technosphere will never have an exergy output equal to its exergy input, as there will always be some loss. For a technosphere with annual exergy output $X$, the exergetic efficiency based on stellar luminosity as a source is $\eta = X/S$, where $S$ and $X$ are evaluated at the same domain radius. A luminosity-limited technosphere is constrained by the limit $\eta < 1$. This constraint states that a luminosity-limited technosphere can never exist exactly at the luminosity limit but will always experience some inefficiencies that cause exergy loss (or, equivalently, increase entropy). The theoretical limits of the Type I, II, III, and IV civilizations on Figs. 1 and 2 may therefore be unattainable for luminosity-limited technospheres.

The second law of thermodynamics also limits the duration at which a technosphere could sustain exponential growth in energy consumption, as the generation of excessive thermal energy as waste would exert direct heating on the planet [e.g.,5,6,20]. The study by Balbi and Lingam [20] estimated that a sustained 1% rate in energy growth from today would cause the loss of habitable conditions on Earth on a timescale of ~1000 years. The consequences of this direct heating are regardless of the source of energy, as all energy consumption will yield thermal waste that will inevitably warm the environment. The problem of direct heating similarly constrains the ability for a civilization to reach the theoretical limit of a Type I civilization, which likewise may also drive such a civilization to expand its spatial scale to better dissipate its generation of thermal waste energy.

It is important to note that $S$ and $X$ are *annual* quantities, tabulated in units of energy (Joules) per unit time (year) in Tables 1 and 2, which is an energy flow over time or *power*. The generation of stellar luminosity itself is the result of a metabolic-like process: stars release luminous energy due to the loss of mass that occurs from the fusion of lighter elements into heavier ones. The total amount of stellar matter lost in this way is very small compared to the mass of the star, so the luminous energy generated over the lifetime of a star is only a small fraction of the total energy that could be available if (hypothetically) the entire stellar mass were converted into luminous energy. A star like the sun has a typical main sequence lifetime of about 10 billion years, which will be followed by expansion into a red giant and eventual evolution into a smaller white dwarf. On this 10 billion year time scale (and even longer), luminous energy remains an energy flow over time (power) that is generated by stellar processes ultimately dependent on the mass of the star. The uppermost limit to exergy in a system is therefore determined by the stellar mass available within the system.

The maximum energy available from stellar mass is $E = c^2 \Sigma M$, where $\Sigma M$ is the total of all mass within the domain radius; this corresponds to the theoretical maximum exergy input based on stellar mass. This is shown as the "solar mass limit" on Fig. 2, which assumes a total of one solar mass within the solar system and $10^{22}$ solar masses for the galaxy. Two of the stationary stellivores reside above the solar mass limit (the stationary accreting neutron star and black hole cases), which indicates that these cases have an even greater metabolic rate that would require a 100 solar mass or larger host to sustain its annual energy use. The maximum energy available from stellar mass is fixed in time (we will limit our consideration to the main sequence lifetime of a star), so the exergetic efficiency, $\epsilon$, based on stellar mass as a source is

$$\epsilon = \frac{Xt}{E} < 1, \qquad (2)$$

where $E$ and $X$ are evaluated at the same domain radius. The annual exergy output for a particular scenario, $X$, is an energy flow in time, so Eq. (2) multiplies $X$ by a specified timescale, $t$. The values of $\epsilon$ relate the exergy outputs of a civilization over a duration of time to the maximum exergy inputs available from stellar mass. Stellivores or any other system that harvests energy from stellar mass may have exergy outputs above the luminosity-based exergetic efficiency (i.e., $\eta > 1$), but such systems remain bound by mass limits (i.e., $\epsilon < 1$). If harvesting mass has greater exergetic efficiency than harvesting radiation (i.e., if $\epsilon > \eta$), then an expanding technosphere may eventually choose to harness more thermodynamically efficient means of utilizing stellar resources.

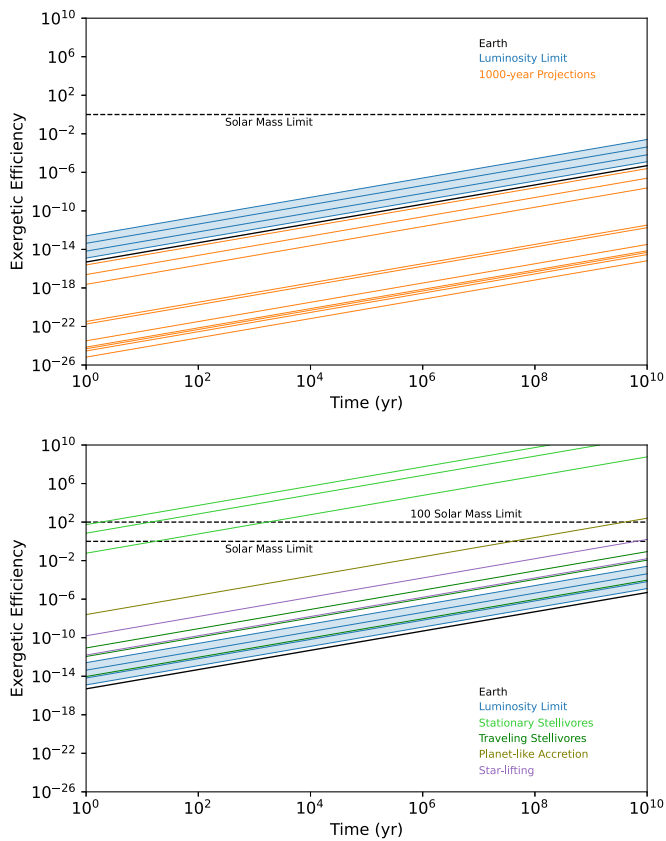
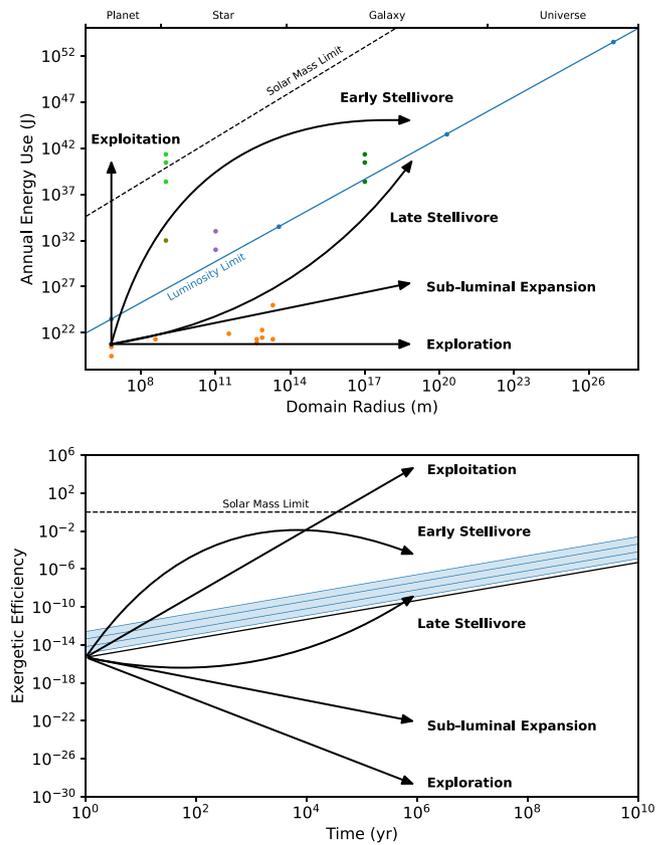

**Fig. 3.** Log–log plot of exergetic efficiency over a 10 billion year timescale for the scenarios considered in Table 1 (top panel) and Table 2 (bottom panel). The exergetic efficiency calculation assumes a solar-mass limit is the maximum energy. Two stationary stellivore cases begin above the solar mass limit, and the third exceeds the solar mass limit at some later time; all of these cases would require a much larger exergy input of 100 solar masses or greater. (For interpretation of the references to color in this figure legend, the reader is referred to the web version of this article.)

**Fig. 4.** Theoretical future trajectories starting from Earth today, shown as annual energy use versus spatial domain radius (top panel) and exergetic efficiency over a 10 billion year timescale (bottom panel). Colored dots in the top panel correspond to the scenarios considered in previous figures. The black line and shaded blue region in the bottom panel correspond to a constant exergy output for Earth and the luminosity limit, respectively. (For interpretation of the references to color in this figure legend, the reader is referred to the web version of this article.)

Fig. 3 shows the exergetic efficiency based on stellar mass for all scenarios considered in Tables 1 and 2, plotted over a 10 billion year timescale. These calculations follow Eq. (2) and assume that the exergy outputs remain constant with time for all scenarios. The luminosity limit is shown as a shaded blue region composed of four lines, the top representing Type IV and the bottom Type I; the increasing trend represents the increased fraction of luminous exergy outputs over time in proportion to the available exergy inputs from stellar mass. The trend for Earth is shown in black, falling below the Type I line. All ten of the 1000-year projections (yellow lines) have a lower exergetic efficiency than present-day Earth, which results from the larger domain radii of these ten scenarios compared to present-day Earth. The traveling stellivore cases all fall beneath the solar mass limit, with one case above the luminosity limit. The star-lifting scenario reaches the solar mass limit by the end of the 10 billion year timescale, which is appropriate as this is a typical main sequence lifetime for a sun-like star. The stationary stellivore cases as well as the planet-like accretion cases all eventually exceed the solar mass limit and even the 100 solar mass limit. These cases would require much higher host star densities, such as the equivalent of a neutron star or black hole, in order to sustain the required mass accretion rates.

This framework provides a way to conceptualize different trajectories that could lead a civilization from conditions like present-day Earth to a stellivore. The possibility space for a civilization's annual energy use and domain radius, as shown in Figs. 1 and 2, involves a trade-off for a given scenario between *exploration* by increasing the domain radius and *exploitation* by increasing annual energy use [21]. A trajectory of pure exploration would involve continual expansion of the domain radius with energy use remaining constant, while a trajectory of pure exploitation would involve a fixed domain radius but a continually increasing energy use. These two extreme cases are shown as horizontal and vertical vectors in the top panel of Fig. 4, along with several other cases. The sub-luminal expansion case involves both an expansion of domain radius and increase in energy use, but at a rate that keeps the civilization below the luminosity limit. The late stellivore case involves a gradually accelerating rate in the increase of energy use, in which the civilization eventually approaches and may even exceed the luminosity limit. The early stellivore case features rapid increases in energy use before the domain radius expands significantly, in which the civilization crosses the luminosity limit much earlier. The late and early stellivore scenarios suggest two trajectories between Earth today and a traveling stellivore, only one of which involves a stationary stellivore as an intermediary state.

The exergetic efficiency of these five trajectories is plotted in the bottom panel of Fig. 4, which assumes that the timescale for the trajectories shown in the top panel is one million years. The exploration and sub-luminal expansion trajectories both show a continual decrease in exergetic efficiency with time, as these involve expansion to galactic scales with reduced or no change to exergetic output. The exploitation trajectory shows a rapid increase in exergetic efficiency, as domain radius remains constant but exergetic outputs increase. The late stellivore trajectory shows an initial decrease followed by an increase in exergetic efficiency and is approaching the luminosity limit. The early stellivore trajectory crosses the luminosity limit rapidly and approaches the solar mass limit before exergetic efficiency begins to reach a limit or decrease. These five possibilities suggest that different evolutionary options may be available for civilizations that increase their exergetic efficiency with time compared to those for which exergetic efficiency decreases.

## 5. Discussion

All of the 1000-year future cases show a lower exergetic efficiency than Earth today (Fig. 3, top panel) because exergetic efficiency is calculated as a function of spatial domain, and all these scenarios expand in spatial domain more rapidly than they increase exergy outputs. This indicates that these scenarios all emphasize an exploration strategy over exploitation [21], which is comparable to the exploration and sub-luminal expansion trajectories (Fig. 4). Two of these scenarios (S4 and S7 in Table 1) have minimized their technospheres within Earth's boundaries, which indicates a focus in these scenarios on purposefully pursuing other energy use options that are less energetically efficient; that is, these cases involve an ideological prioritization to lower exergy outputs over other aspects of their material culture. Other scenarios involve an expansion of the technosphere across the solar system (S9 and S10 in Table 1), but these still prioritize exploration over exploitation and show a decrease in exergetic efficiency as the domain radius increases.

The observation that these 1000-year scenarios all show a decrease in exergetic efficiency may suggest a possible bias underlying these projection methods or reveal underlying assumptions about expectations for future trajectories. In particular, this is because all ten scenarios feature extant and non-speciated humans with the Earth's biosphere still present in some form. In other words, these scenarios all need to proceed with exploration first in order to acquire necessary knowledge to ensure that the waste energy and waste material (and entropy-related degradation in general) is displaced somewhere else in the domain; otherwise, the civilizations in these scenarios would overwhelm their spatial domains with waste or similarly deplete or degrade them. It is possible that following an exploitation trajectory immediately might destroy the host planet's biosphere, so such a trajectory might be viable only after establishing an uninhabited industrial "service world" [22]— for example on Mercury, in the case of the Earth's biosphere, so as not to risk runaway climate change or any other form of degradation that could lead to a biospheric collapse. This problem is related to unavoidable degradation due to absolute entropy increases in smaller domains (including on Earth today), which likewise may be harmful.

The early stellivore trajectory (Fig. 4) involves a transition from Earth today through a planetary accretion phase, followed by a stationary stellivore, and then a traveling stellivore. We can imagine a generalized narrative that might explain such a trajectory. In phase 1, the exponential energy demands of computing leads to a gradual heating up of Earth, eventually transforming it into a fusion powered (post)planet. This is a temporary solution, as the mass of a planet is tiny in comparison to a star, so even using almost all the atoms of Earth as fuel for fusion reactors would eventually run out much more quickly than tapping directly into the host star. Phase 2 would reach the point of planetary accretion or star-lifting, as these strategies support each other: a civilization could extract mass from the host star, thereby prolonging its lifetime, and then use this mass as a new kind of energy source beyond stellar radiation (i.e., beyond the luminosity limit). At phase 3, the civilization begins the transition to a traveling stellivore, as the host star will eventually run out and so another nearby star will be needed. This leads to phase 4, in which the civilization begins accreting another star, and the cycle repeats, with the civilization even pursuing "hardware density transitions" to expand beyond main sequence-type stars to instead sustain metabolism on denser white dwarf, neutron star, or black hole hosts. The extent to which a civilization could persist as a stellivore, particularly one that feeds on large amounts of mass, depends on the availability of sufficient sources of accretable mass to sustain the necessary metabolic rate. With Earth today as a starting point, this early stellivore trajectory would require some form of complete unification and centralization in order to sacrifice the biosphere for the sake of a merged-transhumanity, with a singular vast exergy flow that sustains the entire civilization. By way of metaphor, the early stellivore trajectory is akin to a "hunter–gatherer" approach toward resource collection; alternatively, this is analogous to the ecology of predation, namely operating like true predators or parasitoids depending on how fast they kill their host star.

The late stellivore trajectory (Fig. 4) does not include a planetary accretion or a stationary stellivore phase, but instead this trajectory remains below the luminosity limit until the civilization's domain radius has reached galactic scales. The exergetic efficiency for the late stellivore trajectory begins with a decrease, comparable with the sub-luminal expansion trajectory, but then begins to increase and approach the luminosoty limit. This suggests an alternative path toward becoming a traveling stellivore. One possibility could be that some of the 1000-year scenarios could be consistent with the early stages of a civilization that later becomes a traveling stellivore. Another possibility is that none of these 1000-year scenarios have any logical continuity with a future as a traveling stellivore, but that some other late stellivore trajectory might exist that begins with Earth today. Still another possibility is that no viable late stellivore trajectory exists that includes Earth as a starting point. By way of metaphor, the late stellivore trajectory can be compared to the approach of a "farmer" toward resource acquisition, or from an ecology of predation viewpoint, a grazer or a parasite which does not completely kill its prey (in this case, the host star). The late stellivore trajectory may not necessarily require centralization and unification and instead might involve multiple instances or "backup nodes" of a technosphere that is farming exergy across its domain. Such efforts would attempt to ensure that "recovery copies" of the technosphere exist in multiple places (i.e., different host stars) before attempting to cross the luminosity limit at any one of these stars to create a traveling stellivore, while the distributed civilization can continue to utilize the existing infrastructure around other stars that is already in place. If the domain of a single successfully created traveling stellivore is the entire galaxy, thanks to its precursor civilization, then even when it crosses the luminosity limit of a single host star, its overall place in Fig. 4 (top panel) would be below the luminosity limit point for the whole galaxy—at least until it eventually takes advantage of the infrastructure its civilization has distributed around other stars and starts feeding on them.

Known accreting binary star systems all provide possible examples of early stellivores or traveling stellivores. In this sense, numerous stellivore candidates have already been identified, but the challenge is to find ways to discern between behavior in these systems that would be the result of extraterrestrial activity rather than astrophysical processes. Testing the stellivore hypothesis therefore does not require identifying any new sources (as is the case for other technosignature search efforts) but instead calls for innovative thinking in how to characterize the properties of known accreting binary systems. This could include examining candidate systems for evidence of their utilization as a stellar engine [11], examining the trajectories of such systems for evidence of goal-directed behavior [10], examining their use as an accurate galactic navigation system [23], improving characterization of accretion in such systems to determine if there are any controls that could not be explained through astrophysical processes alone, and applying known biological scaling laws to candidate systems to look for possible evidence of metabolic processes [e.g.,12]. Other possibilities may exist, and future research should continue to explore a wide range of possibilities for examining ways to test the stellivore hypothesis through improved characterization of known accreting binary star systems.

## 6. Conclusion

The exploitation–exploration framework can provide useful insights into thinking about the possible trajectories of Earth's future and advanced technospheres, even very long-lived technospheres that are vastly different than our own. The 1000-year future scenarios illustrate cases closer to pure exploration (including sub-luminal expansion) that are consistent with plausible projections of Earth's technosphere today, given that humans have not speciated in any sense. By contrast, both the early and late stellivore trajectories represent versions of the multi-phased optimal solution to the exploitation–exploration trade-off, depending on the life-stage of a technosphere and its environmental conditions. The framework developed by Berger-Tal et al. [21] (see, e.g., their Figs. 1 and 2) suggests four phases of establishment, accumulation, maintenance, and exploitation with varying durations; however, not all four of these phases are required, and some particular versions of the optimal solution could even involve only the two phases of establishment and exploitation for subjects with very short life-spans. It is worth considering the extent to which the trajectories of advanced or exotic technospheres in particular environments might follow different versions of such a multi-phased solution that optimizes their exploration–exploitation strategies in ways that might be inaccessible to Earth's technosphere today.

Ultimately, any hypothetical exploration of civilizational trajectories, no matter how exotic or mundane, should lead to observational predictions if they are to be useful in aiding the search for technosignatures. One important lesson from this study is to recognize that luminosity and mass both serve as physical limits to growth, and a civilization's strategy in maximizing its expansion in spatial scale versus its increases in energy use can lead to different observable consequences. The 1000-year scenarios that fall close to the pure exploration trajectory may prove to be among the most difficult to detect—or, at least, discovering such a civilization may require more serendipitous conditions when compared to others that expand in both spatial scale and energy use. But the trajectories that are closer toward an optimal balance between exploration and exploitation also raise challenges in observation, given that such technospheres may be vastly different than that on Earth, and so recognizing their associated technosignatures may not be obvious. For stellivores, possible approaches toward testing this hypothesis include examining known accreting binary systems as candidate traveling stellivores for possible signatures of a stellar engine [11], testing the goal-directedness of candidate traveling stellivore [10], testing whether accretion in candidate stationary stellivores is controlled, and applying biological metabolic scaling laws to candidate stellivores [e.g.,12]. Such efforts would be an important first step toward understanding the extent to which known stellar systems could reveal unusual properties that are closer to living systems than they first appear.

In summary, this paper calls for more broad and imaginative thinking about the possible trajectories that could lead to a remotely detectable technosphere. Previous attempts at searching for characteristic signs of Type I, II, III, and IV civilizations may inadvertently presume that an advanced civilization will develop along the luminosity limit as a trajectory, but achieving the luminosity limit itself remains out of bounds for a luminosity-limited technosphere due to thermodynamic constraints. Likewise, any civilization that can exist at or beyond the luminosity limit will necessarily have discovered a way to derive greater exergetic efficiency from its environment, such as by harvesting stellar mass directly instead of relying only on stellar luminosity.

## CRediT authorship contribution statement

**Jacob Haqq-Misra:** Writing – original draft, Conceptualization. **Clément Vidal:** Writing – review & editing, Conceptualization. **George Profitiliotis:** Writing – review & editing, Conceptualization.

## Declaration of competing interest

The authors declare that they have no known competing financial interests or personal relationships that could have appeared to influence the work reported in this paper.

## Acknowledgments

This article was inspired by the session on "Advancing the Search for Technosignatures" and other conversations at the 2024 Astrobiology Science Conference in Providence, Rhode Island. J.H.M. gratefully acknowledges support from the NASA Exobiology program under grant 80NSSC22K1009. Any opinions, findings, and conclusions or recommendations expressed in this material are those of the authors and do not necessarily reflect the views of their employers or NASA.